\documentclass[3p,times]{elsarticle}

\usepackage{ecrc}

\usepackage{epstopdf}

\volume{00}

\firstpage{1}

\journalname{Nuclear Physics A}

\runauth{A. Marin et al.}

\jid{nupha}

\jnltitlelogo{Nuclear Physics A}

\usepackage{graphicx}
\usepackage{amsmath,amssymb}

\begin{document}

\begin{frontmatter}



\title{Neutral meson production in
pp and Pb-Pb collisions
measured by ALICE
at the LHC}

\author[label1,label2]{A. Mar\'{\i}n (for the ALICE\fnref{col1} Collaboration)}
\fntext[col1] {A list of members of the ALICE Collaboration and acknowledgements can be found at the end of this issue.}

\address[label1]{Research Division and ExtreMe Matter Institute EMMI}
\address[label2]{GSI Helmholtzzentrum f\"ur Schwerionenforschung, Darmstadt, Germany,}




\begin{abstract}
The midrapidity $\pi^0$ nuclear modification factor, $R_{\rm{AA}}$,  at $\sqrt{s_\mathrm{NN}} =$ 2.76 TeV  in 6 centrality 
classes as well as the corresponding $\pi^0$  invariant yields in Pb-Pb and in pp collisions are presented.
The transverse momentum range covered is 0.6 (0.4) GeV/$c$ $< p_{\mathrm{T}} <$ 12 (10) GeV/$c$ for Pb-Pb (pp) collisions. 
A suppression of neutral pions increasing with centrality is observed. The yield of charged particles associated with a 
high $p_{\mathrm{T}}$ neutral pion trigger  
(8 GeV/$c$ $< p_{\mathrm{T}} <$ 16 GeV/$c$) is measured in pp and Pb-Pb collisions at 
$\sqrt{s_\mathrm{NN}} =$ 2.76 TeV. The conditional per-trigger yield modification factor in the near and away side is in agreement with the
measured one for charged particles.
\end{abstract}

\begin{keyword}
Hadron production \sep heavy-ion collisions \sep LHC

\end{keyword}

\end{frontmatter}



\vspace{1cm}
Since the startup of the CERN LHC data from pp collisions at $\sqrt{s} =$~0.9, 2.76, 7 and 8~TeV and Pb-Pb collisions at $\sqrt{s_\mathrm{NN}} =$ 2.76 TeV are available.  An energy density of $\epsilon \sim$ 15 GeV/fm$^3$ at a collision time of 1 fm/$c$ \cite{Collaboration:2011rta} (about 3 times larger than at RHIC)
is reached in Pb-Pb collisions that is above the critical energy density of 1 GeV/fm$^3$ to undergo the predicted QCD phase transition to the Quark-Gluon Plasma (QGP, deconfined state of matter). 
The energy loss of hard partons produced, in the early stage of the collisions, via radiative or collisional mechanisms when traversing the QGP 
translates into modifications of the high $p_\mathrm{T}$ hadron yields in Pb-Pb collisions when compared to pp collisions. This effect first observed at RHIC for charged particles and neutral pions \cite{Adcox:2001jp} has been also reported at the CERN LHC \cite{Aamodt:2010jd}. In order to quantify the suppression the nuclear modification factor is computed from the measurement in Pb-Pb and pp collisions
using: 
\begin{equation}
R_{\rm{AA}}(p_\mathrm{T}) = {{\mathrm{d}^2N/dp_\mathrm{T}\mathrm{d}y|_{\rm{AA}}}\over{\langle T_{\rm{AA}}\rangle\mathrm{d}^2\sigma/\mathrm{d}p_\mathrm{T}\mathrm{d}y|_\mathrm{pp}}}
\end{equation}
where $\langle T_{\rm{AA}}\rangle=\langle N_\mathrm{coll}\rangle/\sigma^\mathrm{pp}_\mathrm{inel}$ and the number of binary nucleon-nucleon collisions 
$\langle N_\mathrm{coll}\rangle$ is obtained from a Glauber model. The neutral as well as the charged pion measurement allows to investigate the differences 
on the suppression pattern between mesons and baryons. 
Moreover, the production of neutral pions at LHC energies is dominated  by gluon fragmentation \cite{Sassot:2010bh}, contrary to RHIC where the contribution
from quark fragmentation is important \cite{Sassot:2009sh}. 
Therefore, the measurement of neutral pions provides constraints on the gluon to pion fragmentation function.
In addition, the comparison of the spectra to theoretical calculations for pp and Pb-Pb collisions
can help to understand the particle production and the characteristics of the QGP formed in heavy-ion collisions.
Finally, the precise measurement of $\pi^0$ and $\eta$ meson spectra over a large $p_{\rm{T}}$ range is a
 prerequisite for understanding the decay photon (electron) background for a direct photon (charm and beauty) measurement.
In this paper, we present the measurement of the $\pi^0$ invariant yield in six centrality classes in Pb-Pb collisions and in pp collisions at the
same energy and the resulting  $R_{\rm{AA}}$.  
The measurement of $\pi^0$-hadron correlations and of the conditional per-trigger yield modification on the near and away side is also presented.

Neutral pions are measured in ALICE in the two photon decay channel in a broad $p_\mathrm{T}$ range \cite{Abelev:2014ffa} using the 
EMCal or PHOS calorimeters or by measuring the $e^+e^-$ pairs produced in  photon conversions in the detector material in the 
central barrel (ITS+TPC). PHOS (EMCal)  is made of a high granularity lead tungstate crystals (large acceptance lead scintillator sandwich) 
and covers $|\eta| < $0.3 ( 0.7) and $\Delta \phi = 60^\circ$ ($100^\circ$). The photon conversion method (PCM) has full azimuthal coverage but the photon conversion probability is only 8.5\% ( $|\eta| < $0.9 ), as ALICE was optimized for a low material thickness.

\begin{figure}[ht]
\begin{flushleft}
\begin{minipage}{0.52\linewidth}
\vspace{-0.25cm}
\includegraphics*[width=0.95\textwidth]{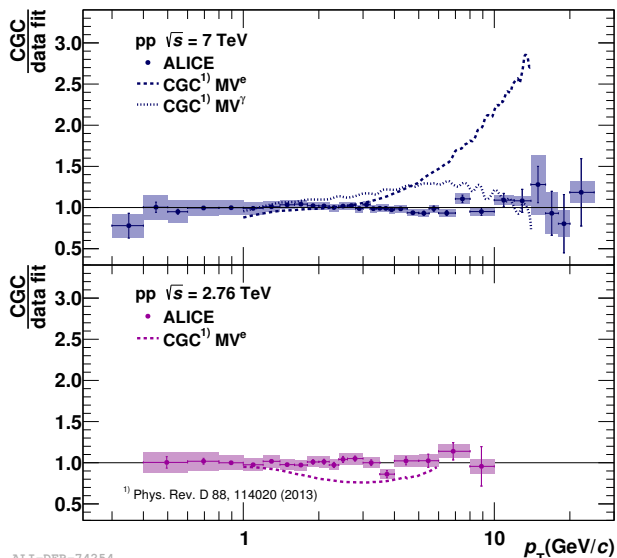}
\end{minipage}
\end{flushleft}
\vspace{-8.75cm}
\begin{flushright}
\begin{minipage}{0.48\linewidth}
The $\pi^0$ invariant yields and $R_{\rm{AA}}$  results are obtained from a sample of 
16.1~$\times$~10$^6$ and 13.2~$\times$~10$^6$ events collected in the 2010 Pb-Pb run and of 34.7~$\times$~10$^6$ 
and 58~$\times$~10$^6$ events
 from the  2011 pp run at $\sqrt{s_\mathrm{NN}}=2.76$~TeV for PHOS  and PCM, respectively \cite{Abelev:2014ypa}. The spectra are the
 weighted average of the PHOS and the PCM results. Next to leading order perturbative QCD predictions computed with the DSS fragmentation function (FF) and the CTEQ6M5 parton distribution function calculations done with the DSS FF overestimate the  
 $\pi^0$  yield in pp collisions at  $\sqrt{s}$ = 2.76, and 7~TeV  \cite{Abelev:2014ypa,Abelev:2012cn} by more than a factor 1.5, while they described the data for $\sqrt{s}$~=~0.9 TeV~\cite{Abelev:2012cn}. 
 On the other hand, the $\pi^0$  production computed within the Color Glass Condensate framework \cite{Lappi:2013zma} describes
the data (see Fig.~\ref{fig:CGCratio}) without any additional factor up to a moderate or a high $p_\mathrm{T}$ with two different initial conditions. 
The parameters of the 
model were fixed to the deep inelastic scattering data. The DSS fragmentation function at LO is used.
Further comparisons between the pQCD and the CGC calculations could reveal 
issues not only on the FF \cite{d'Enterria:2013vba} but also on the initial state gluon distributions.

\end{minipage}
 \end{flushright}
 \vspace{-2.25cm}
\begin{flushleft}
\begin{minipage}{0.48\linewidth}
\caption{Ratio of the CGC calculations to the fit of the $\pi^0$ invariant  cross section or invariant yield as measured by ALICE in pp collisions 
at $\sqrt{s}=$~7~TeV (top) and $\sqrt{s}=$~2.76~TeV (bottom), respectively.}
\label{fig:CGCratio}
\end{minipage}
\end{flushleft}
\end{figure}

\begin{figure}[hp]
\begin{tabular}{lr}
\includegraphics*[width=0.45\textwidth]{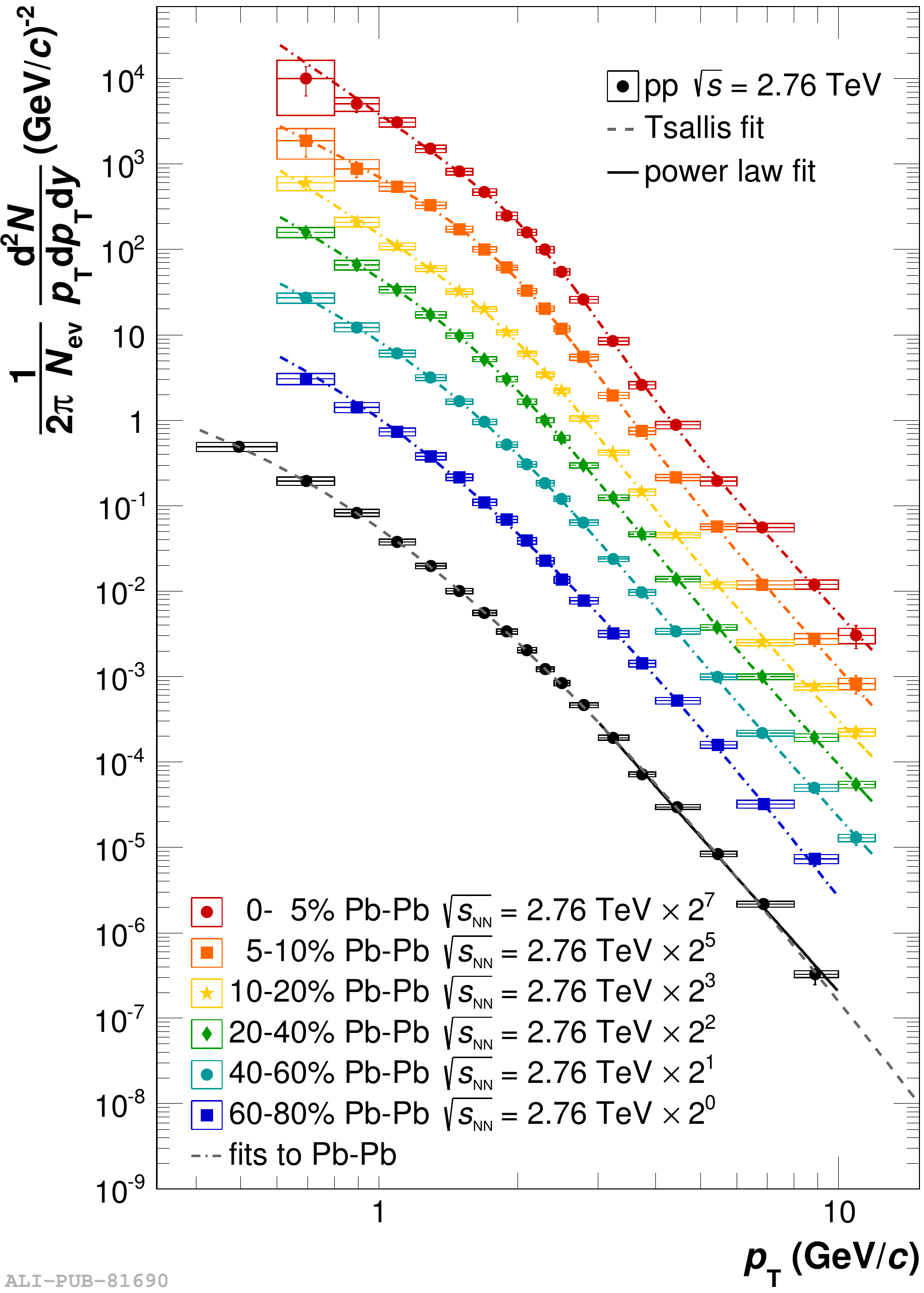}&
\includegraphics*[width=0.45\textwidth]{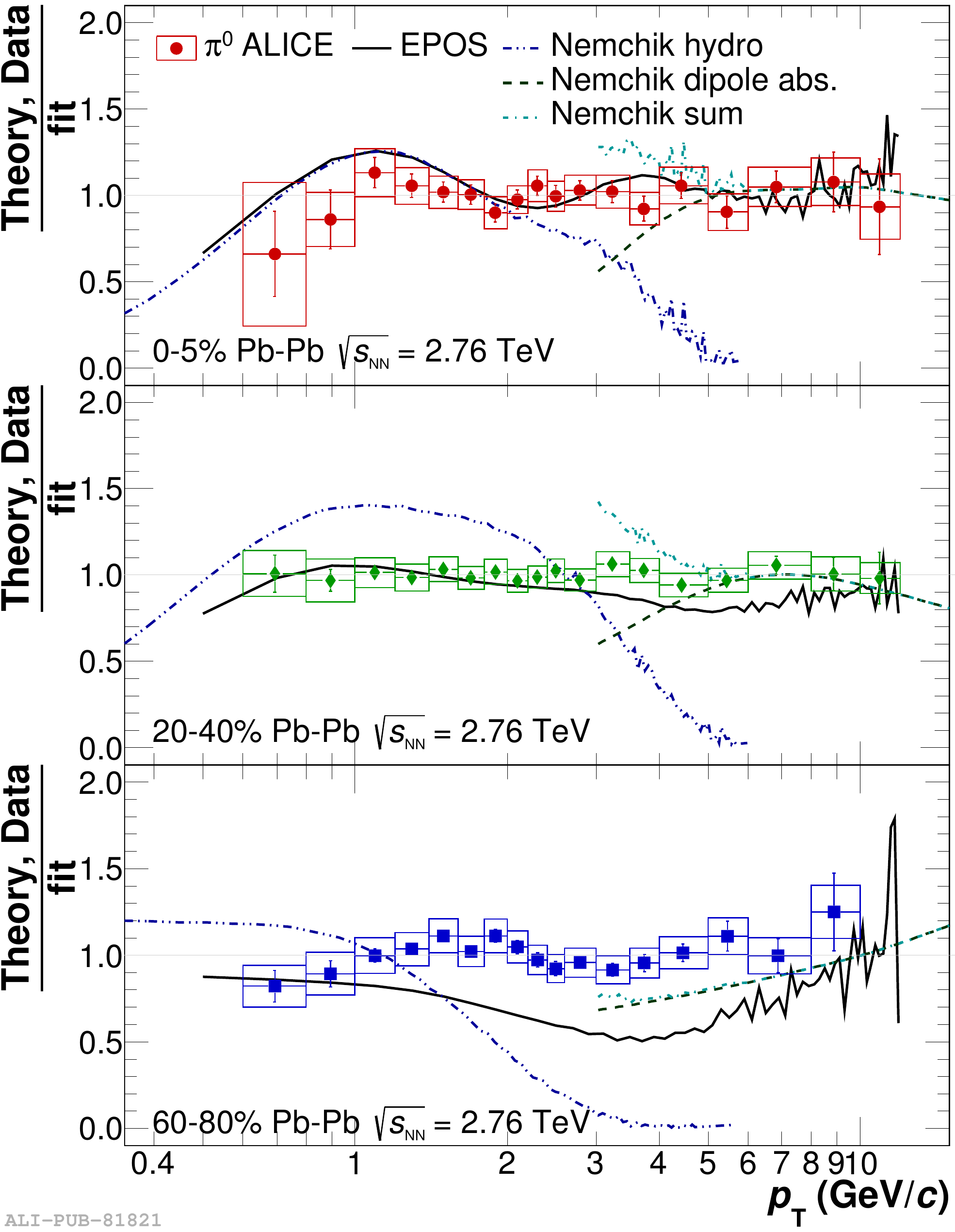}
\end{tabular}
\caption{Left: $\pi^0$ invariant yield in Pb-Pb collisions in 6 centrality classes and in inelastic pp collisions at $\sqrt{s_\mathrm{NN}}~=$~2.76 TeV.  Right: Ratio of the theoretical predictions from the EPOS model and by Nemchick {\it et al.} to the fit of the invariant yield in Pb-Pb collisions for 3 centrality classes.}
\label{fig:pi0PbPb}
\end{figure}

Neutral pions have been measured in Pb-Pb collisions in 6 centrality classes (Fig.~\ref{fig:pi0PbPb}, left).
There is a first attempt of theoretical models, EPOS \cite{Werner:2012xh} and calculations by Nemchick {\it et al.} 
\cite{Nemchik:2013ooa,Kopeliovich:2012sc} to describe the complete $\pi^0$  spectrum in Pb-Pb collisions (see Fig.~\ref{fig:pi0PbPb}, right).
The EPOS model contains hydrodynamical flow at low $p_\mathrm{T}$
 and energy loss of high $p_\mathrm{T}$ string segments at high $p_\mathrm{T}$. It reproduces the spectrum for central and semi-central collisions
 while it develops a discrepancy for peripheral collisions, maybe due to an underestimation of the hydrodynamical flow contribution in 
 the $p_\mathrm{T}$ range between 1 and 5 GeV/$c$. 
\begin{figure}[hp]
\begin{center}
\includegraphics*[width=0.95\textwidth]{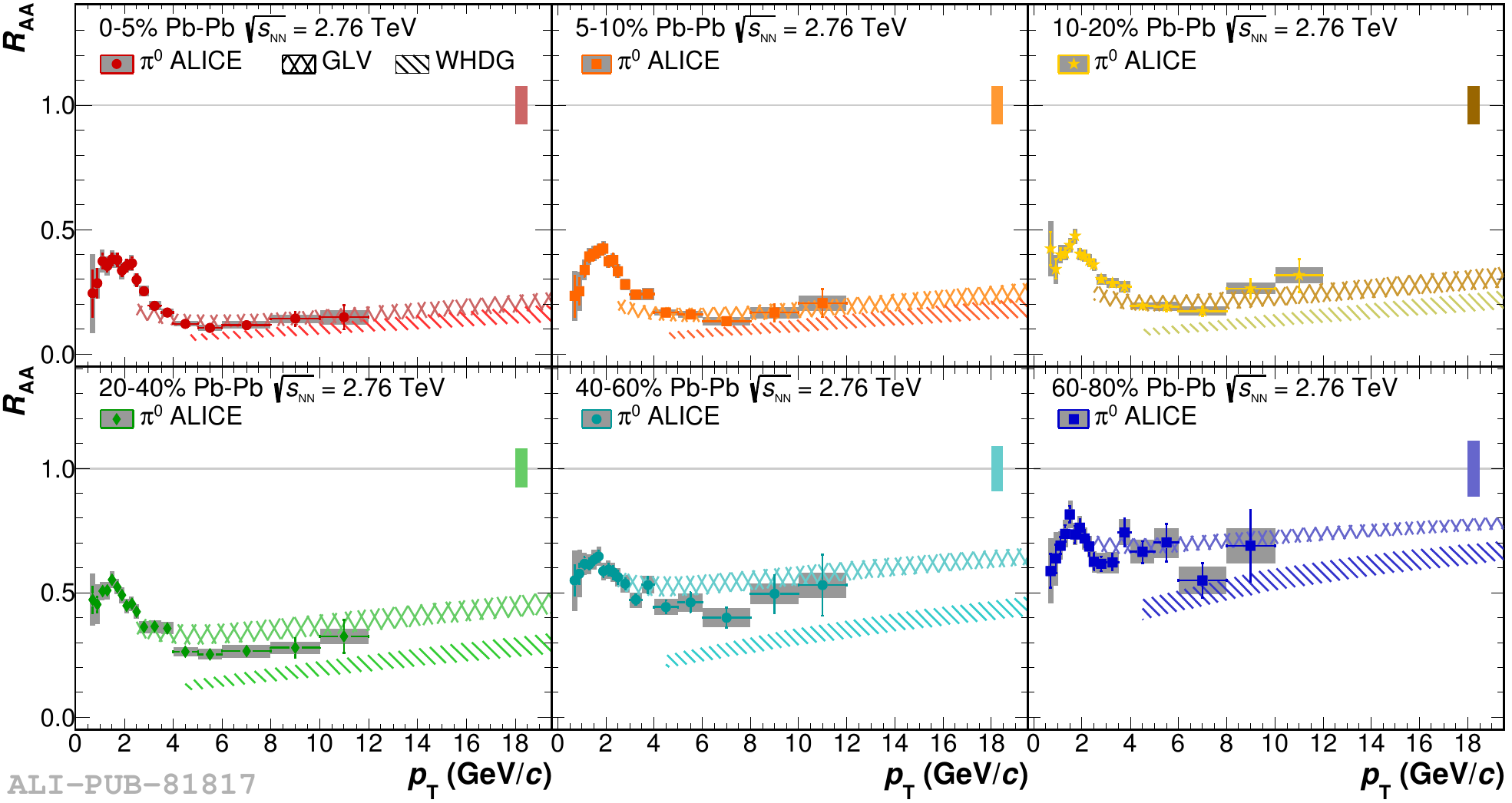}
\caption{Measured $\pi^0$ $R_{\rm{AA}}$ in six centrality classes compared to theoretical predictions from the GLV and the WHDG theoretical models.}
\label{fig:RAA}
\end{center}
\end{figure}
The calculations by Nemchik {\it et al.} contain a hydrodynamical description at low $p_\mathrm{T}$ and color dipole absorption 
at high $p_\mathrm{T}$. They describe the spectrum in central collisions except in the intermediate $p_\mathrm{T}$ region. For 
semi-central collisions the spectrum is overpredicted at low $p_\mathrm{T}$.  This model also predicts other observables like the 
azimuthal asymmetry \cite{Nemchik:2014gka}.

The $\pi^0$ nuclear modification factor $R_{\rm{AA}}$  in six centrality 
classes is shown in Fig.~\ref{fig:RAA}. The $\pi^0$ suppression increases as the centrality increases from $R_{\rm{AA}}=$~0.6 in the 60-80\% centrality class down to $R_{\rm{AA}} =$~0.1 for the 0-5\% centrality class.
 The theoretical predictions from the GLV~\cite{Sharma:2009hn,Neufeld:2010dz}  and WHDG~\cite{Horowitz:2007nq} models that were tuned to reproduce the RHIC data are 
 compared to the experimental results in Fig.~\ref{fig:RAA}.
 The GLV model including final state energy loss and nuclear broadening is able to reproduce the $p_\mathrm{T}$ and centrality dependence of the suppression.
 The WHDG model including collisional and radiative parton energy loss and geometrical path length fluctuations reproduces the $p_\mathrm{T}$ dependence of the $\pi^0$ suppression but overestimates it for more peripheral collisions. The $\pi^0$ $R_{\rm{AA}}$ at the LHC is lower than at SPS and at RHIC.
 The $p_\mathrm{T}$ dependence of the suppression at the LHC is similar to that at RHIC at $\sqrt{s_{NN}} =$~0.2 TeV.

\begin{figure}[hbt]
\begin{center}
\begin{tabular}{lll}
\includegraphics*[width=0.315\textwidth]{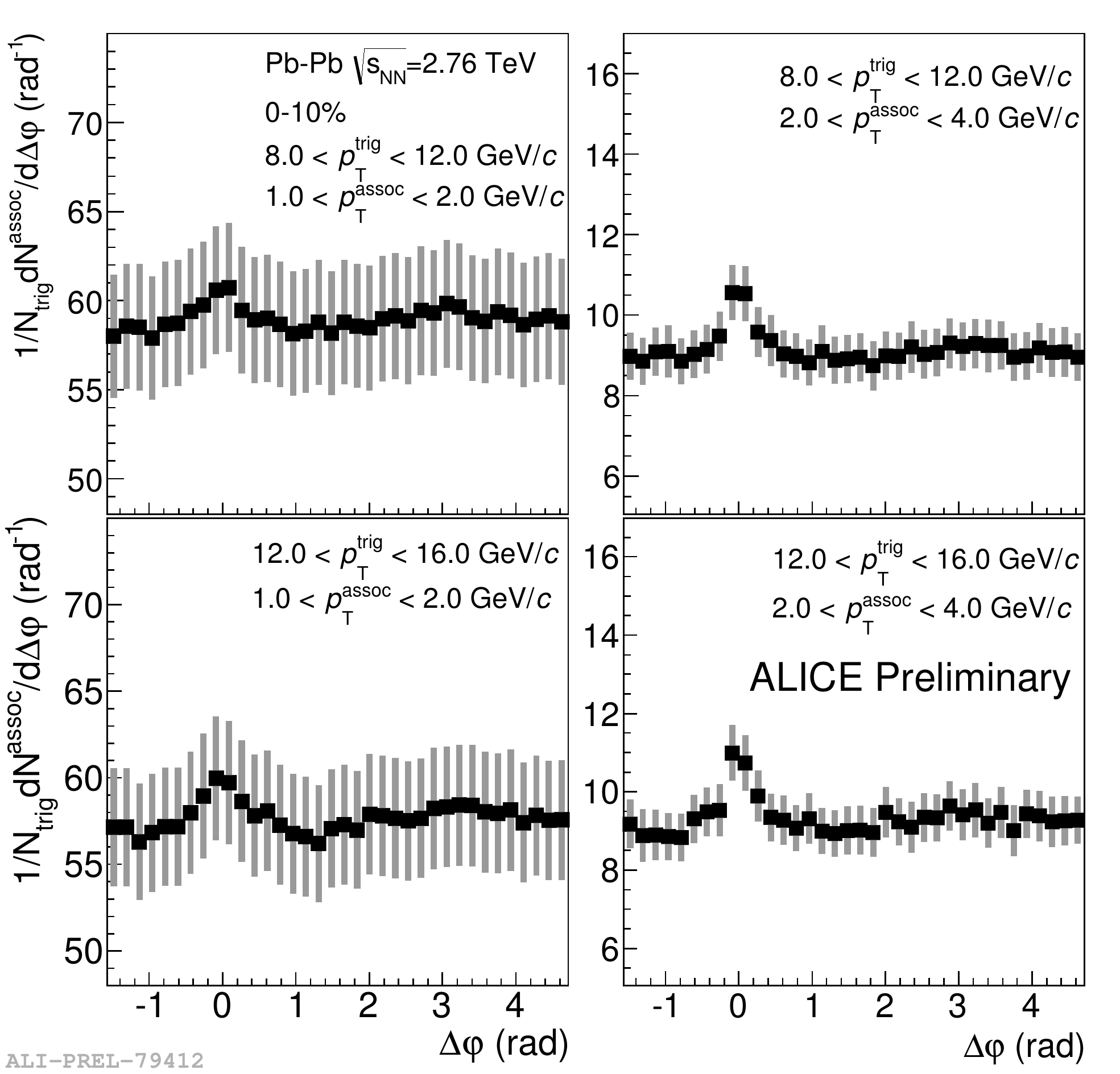}&
\includegraphics*[width=0.315\textwidth]{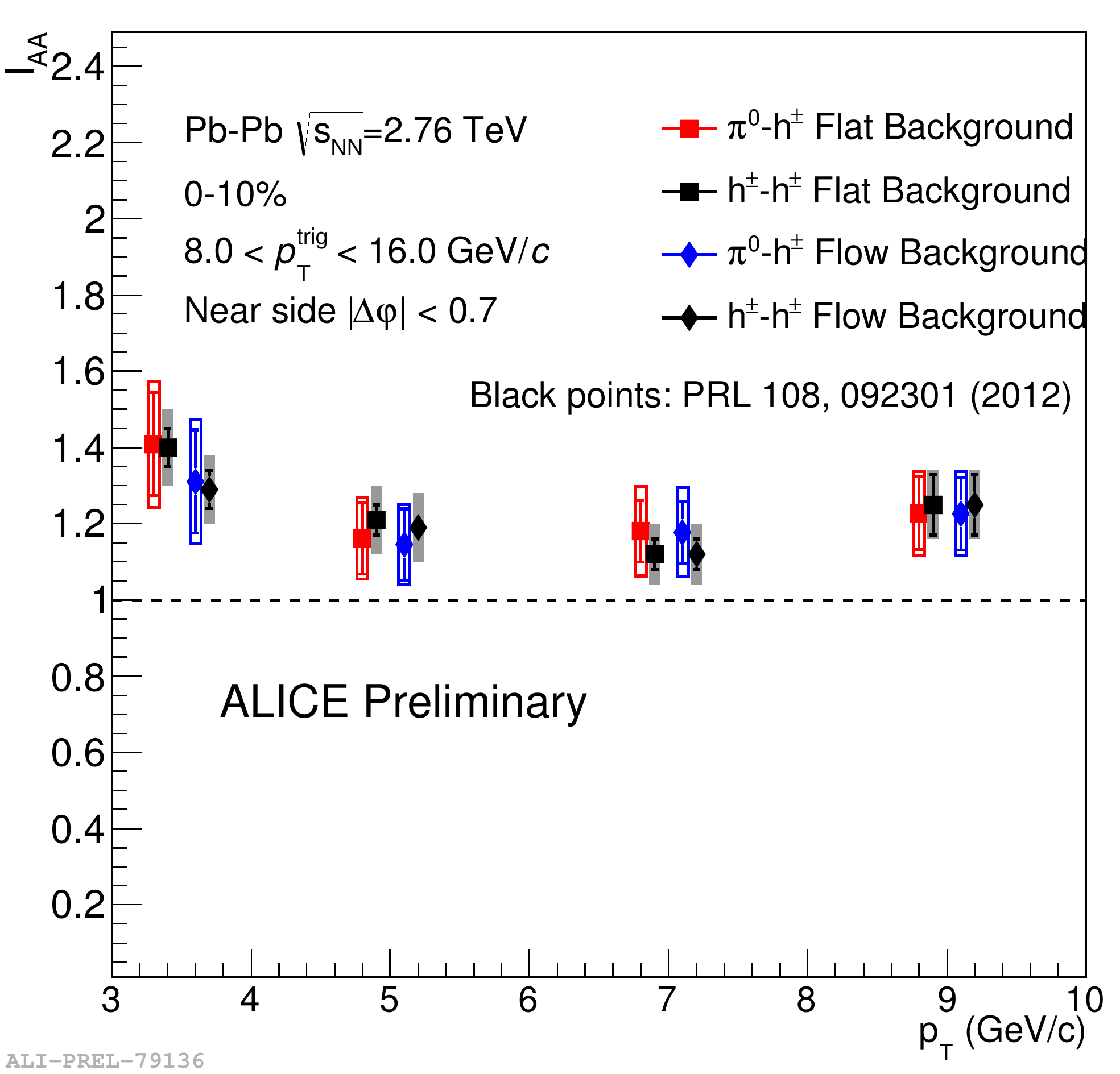}&
\includegraphics*[width=0.315\textwidth]{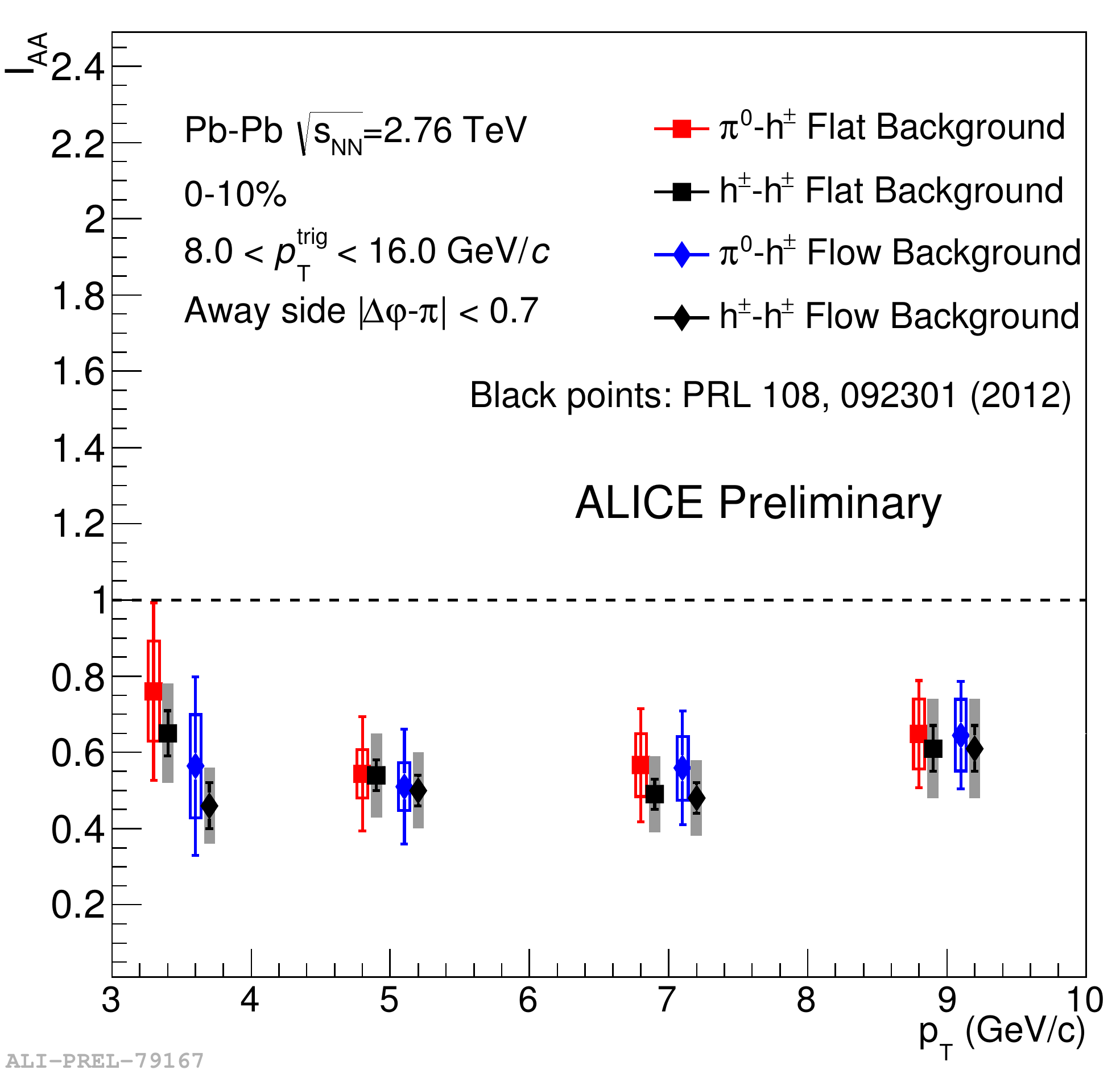}\\
\end{tabular}
\caption{Left: $\pi^0$-hadron correlations for 0-10\% central Pb-Pb collisions for different trigger and associated $p_\mathrm{T}$ with 
corrections for contamination, resolution and efficiency included. Middle:
Conditional per-trigger yield modification factor as function of the associated $p_\mathrm{T}$ in the near side compared to the 
results from unidentified charged particles.
Right: Conditional per-trigger yield modification factor as function of the associated $p_\mathrm{T}$ in the away side compared to the results from unidentified charged particles.}
\label{fig:IAAA}
\end{center}
\end{figure}

 In  order to get further insight into the jet quenching mechanism and medium induced
parton energy loss in the QGP,  $\pi^0$- hadron correlations have been studied using a sample of 0.63~$\times$~10$^6$ EMCal triggered 
events (13.6 nb$^{-1}$ integrated luminosity) and 16.5 ~$\times$~10$^6$ events in the 0-10\% centrality class from the 2011 run in pp and Pb-Pb collisions at $\sqrt{s_{NN}}=~$2.76 TeV, respectively  (Fig.~\ref{fig:IAAA}, left).  
Neutral pions are detected
  in the EMCal and the charged particles in the central barrel.
Azimuthal correlations are a more differential observable than $R_{\rm{AA}}$ that carry information about the jet shape modification by comparing the less affected trigger jet and the more affected away side jet.
 The modification factor of conditional per-trigger yield of charged hadrons is computed using
 \begin{equation}
I_{\rm{AA}} (p_{T}^{\pi^0},p_{T}^{h^\pm}) = {{Y^{PbPb}(p_{T}^{\pi^0},p_{T}^{h^\pm}) }\over{Y^{pp}(p_{T}^{\pi^0},p_{T}^{h^\pm}) }}
 \end{equation}
 A factor $I_{\rm{AA}}\sim$~1.2 enhancement versus the associated  $p_\mathrm{T} $ is visible on the near side (Fig.~\ref{fig:IAAA}, middle) that was not observed 
 at lower energies. A suppression of a factor $I_{\rm{AA}}\sim$~0.6 on the 
 away side (Fig.~\ref{fig:IAAA}, right) at $p_\mathrm{T} >$~3 GeV/$c$ is an evidence for the in medium energy loss.
The results are consistent with the ones obtained for charged particles~\cite{Aamodt:2011vg}.

We would like to thank W. Vogelsang for providing the NLO QCD calculations used in this paper
and T. Lappi and H. M\"antysaari for providing the CGC calculations.

\end{document}